\documentclass[pdflatex,sn-mathphys-num]{sn-jnl}

\usepackage{graphicx}%
\usepackage{multirow}%
\usepackage{amsmath,amssymb,amsfonts}%
\usepackage{amsthm}%
\usepackage{mathrsfs}%
\usepackage[title]{appendix}%
\usepackage{xcolor}%
\usepackage{textcomp}%
\usepackage{manyfoot}%
\usepackage{booktabs}%
\usepackage{algorithm}%
\usepackage{algorithmicx}%
\usepackage{algpseudocode}%
\usepackage{listings}%
\usepackage{geometry}
\theoremstyle{thmstyleone}%
\theoremstyle{thmstyletwo}%

\theoremstyle{thmstylethree}%

\begin{document}

\title[Article Title]{On the Reliability of AI Methods in Drug Discovery: Evaluation of Boltz-2 for Structure and Binding Affinity Prediction}


\author*[1]{\fnm{Shunzhou} \sur{Wan}}\email{shunzhou.wan@ucl.ac.uk}
\equalcont{These authors contributed equally to this work.}

\author[1]{\fnm{Xibei} \sur{Zhang}}
\equalcont{These authors contributed equally to this work.}

\author[1]{\fnm{Xiao} \sur{Xue}}

\author*[1,2]{\fnm{Peter V.} \sur{Coveney}}\email{p.v.coveney@ucl.ac.uk}

\affil[1]{\orgdiv{Centre for Computational Science}, \orgname{University College London}, \orgaddress{\city{London}, \postcode{WC1H 0AJ}, \country{U.K.}}}

\affil[2]{\orgdiv{Advanced Research Computing Centre}, \orgname{University College London}, \orgaddress{\city{London}, \postcode{WC1H 0AJ}, \country{U.K.}}}



\abstract{Despite continuing hype about the  role of AI in drug discovery, no ``AI-discovered drugs'' have so far received regulatory approval. Here we assess one of the latest AI based tools in this domain. The ability to rapidly predict protein-ligand structures and binding affinities is pivotal for accelerating drug discovery. Boltz-2, a recently developed biomolecular foundation model, aims to bridge the gap between AI efficiency and physics-based precision through a joint ``co-folding'' approach. In this study, we provide an extensive evaluation of Boltz-2 using two large-scale datasets: 16,780 compounds for 3CLPro and 21,702 compounds for TNKS2. We compare Boltz-2 predicted structures with traditional docking and binding affinities with binding free energies derived from the physics-based ESMACS protocol. Structural analysis reveals significant global RMSD variations, indicating that Boltz-2 predicts multiple protein conformations and ligand binding positions rather than a single converged pose. Energetic evaluations exhibit only weak to moderate correlations across the global datasets. Furthermore, a focused analysis of the top 100 compounds yields no significant correlation between the Boltz-2 predictions and the binding free energies from fine-grained ESMACS, alongside observed saturation difference in ligand structures. Our results show that while Boltz-2 offers substantial speed for initial screening, it lacks the energetic resolution required for lead identification. These findings highlight the necessity of employing physics-based methods for the reliability and refinement of AI-derived models.}

\maketitle

\section{Introduction}\label{sec1}

More than a decade has passed since the initial wave of optimistic projections suggested that artificial intelligence (AI) would fundamentally transform the landscape of drug discovery\cite{Lavecchia_2015, Chen_2018}. During this period, a wide range of promising tools and deep-learning architectures have proliferated, claiming the ability to revolutionise every stage of the early drug development pipeline – from the identification of novel drug targets and the prediction of their 3D structures to the rapid identification and optimisation of lead compounds\cite{chen_2023_artificial}. These efforts reached a significant milestone with the development of AlphaFold 2\cite{AlphaFold2_2021}, a breakthrough in the high-fidelity prediction of protein structures that was subsequently recognised with the 2024 Nobel Prize in Chemistry\cite{AlphaFold2_Nobel}. However, despite such unprecedented successes, the impact of AI on drug discovery remains a subject of critical debate. While AlphaFold 2 provides valuable structural hypotheses that can significantly accelerate research, recent evaluations emphasise that its predictions do not replace the necessity of experimental structure determination\cite{terwilliger_alphafold_2024}. Consequently, the expected revolution in the delivery of novel, clinically successful therapeutics has yet to fully materialise\cite{Bender_2021,AI_drugs_2025}. Many AI-driven predictions have failed to translate into reliable experimental outcomes, highlighting the persistent challenge of accurately quantifying the interaction energetics required for hit identification and lead optimisation\cite{masters_co_folding_2025}.

Indeed, the precise determination of the binding affinities of small molecules to their target proteins remains a cornerstone of structure-based drug design. This capability directly dictates the efficiency of the entire discovery pipeline, as the precision of these energetic estimations is essential for the effective identification and subsequent optimisation of novel drug candidates\cite{sadybekov2023}. Traditionally, the field has relied on a trade-off between high-throughput, lower-precision methods like molecular docking\cite{docking} and computationally demanding, high-precision physics-based approaches. The latter approaches include linear interaction energy (LIE)\cite{LIE_1998}, molecular mechanics energies combined with Poisson-Boltzmann or generalized Born and surface area continuum solvation (MMPBSA and MMGBSA) methods\cite{MMPBSA_2015} and alchemical methods\cite{Chodera_2011}, which have been broadly applied to model molecular recognition for drug discovery and lead optimization (an example of the application of alchemical methods can be found in the literature\cite{Schindler2020}).

Standard MMPBSA or alchemical free energy calculations typically rely on one-off simulations, which can be highly sensitive to initial conditions and fail to sample the conformational landscape adequately, leading to significant statistical uncertainties and poor reproducibility\cite{Coveney_2016,Coveney_Wan_2025}. To bridge the gap between theoretical precision and practical reliability, we have developed ensemble-based protocols such as ESMACS (enhanced sampling of molecular dynamics with approximation of continuum solvent)\cite{ESMACS_2015,Wan_2020} and TIES (thermodynamic integration with enhanced sampling)\cite{TIES_2017,Wan_2020}. These methods build upon the foundations of MMPBSA and thermodynamic integration by incorporating an ensemble approach to rigorously address the inherent stochastic properties of molecular dynamics (MD) simulations\cite{Coveney_2016}. By executing multiple independent replicas with varying initial configurations, ESMACS and TIES, which identify lead molecules and perform lead optimisation, respectively, provide reliable and reproducible binding free energy estimates with quantification of statistical uncertainties. ESMACS has been applied to datasets of tens of thousands of compounds on exascale computers for drug discovery\cite{Loeffler2024}. Although it could be increased tenfold without trouble, the set of compounds is still very small, compared to the vast chemical space that needs to be explored (estimated to be about $10^{68}$ compounds). The high computational cost of physics-based approaches limits their applications to relatively small sets of compounds, necessitating faster alternatives for large-scale screening.

Recent advances in generative artificial intelligence (AI) have substantially expanded the accessible chemical space for drug discovery\cite{schneider2020rethinking}, enabling the rapid design of novel compounds through deep generative models\cite{gomez2018automatic,nigamaugmenting} and reinforcement learning-based optimisation\cite{Loeffler2024,sanchez2018inverse}. The development of surrogate docking approaches\cite{clyde2023docking} makes it practical to rapidly screen a vast number of compounds from existing libraries. In practice, these approaches remain critically dependent on fast surrogate scoring functions, which often lack explicit physical grounding and are liable to introduce systematic biases during optimisation\cite{walters2021critical}.

While these task-specific models often function in an isolated context, there is a growing requirement for more integrated, multi-modal systems that unify diverse biomolecular predictions. The emergence of ``foundation models''\cite{foundation_models_nsr2025} has subsequently shifted the landscape of AI. They are large-scale architectures trained on massive, diverse datasets that can be adapted to a wide range of downstream tasks. These models are intended to serve as a versatile ``foundation'' for many different applications. It should be noted that, like all AI methods, foundation models cannot magically create or predict arbitrary molecules and molecular structures; they are only as good as the data they are trained on.

Building on the requirement for both precision and speed, Boltz-2\cite{Boltz2_2025} has recently emerged as a pioneering structural biology foundation model designed to bridge the gap between AI and rigorous physics. It is categorized as a billion parameter foundation model because it is trained on a vast dataset of protein structures and millions of experimental binding records. While other ``co-folding'' frameworks such as AlphaFold 3\cite{AF3}, RoseTTAFold All-Atom\cite{RFAA} and the initial Boltz-1 release\cite{boltz_1} primarily focus on high-fidelity structural assembly, the unified architecture of Boltz-2 extends these capabilities to include direct, quantitative binding affinity prediction. Notwithstanding the vast array of ML methods that has been specifically designed for this purpose in recent years, achieving high-fidelity accuracy remains a significant challenge\cite{ML_for_BFE_2024}. A key strength of Boltz-2 is its claimed proficiency in ranking analogues within a chemical series, allowing it to distinguish subtle variations in potency for hit-to-lead and lead optimization. Consequently, Boltz-2 represents a potential paradigm shift in lead identification, offering a high-throughput alternative that maintains the structural and energetic nuance required for drug discovery. It is this specific claim of predictive accuracy, purportedly reaching the accuracy levels of free energy perturbation (FEP) while operating orders of magnitude faster\cite{Boltz2_2025}, within a single integrated framework, that motivates the study reported here.  

Since its release, Boltz-2 has attracted much attention in peer-reviewed articles\cite{Bret_2026,Boltz_ABFE}, preprints\cite{Boltz2_Glues_2025,grieswelle_boltz2_2026}, and blogs\cite{Boltz2_blog_ryan, Boltz2_blog_Wagen}, which seek to evaluate the reliability of its predictions and its domains of applicability. In particular, Bret \textit{et al}.\cite{Bret_2026} demonstrated that, although Boltz-2 excels as a binary classifier, it exhibits limited precision when quantitatively ranking compounds by affinity. However, the performance of Boltz-2 across large-scale, target-specific chemical libraries remains to be rigorously validated against established physics-based benchmarks. Furthermore, drug discovery frequently explores complex molecular structures and millions of novel candidates with little or no training data, posing a substantial challenge even for foundation models that perform well on more common molecules. 

In this study, we present a comprehensive evaluation of Boltz-2’s predictive capabilities using two distinct and therapeutically relevant protein targets: the SARS-CoV-2 main protease (3CLPro), a critical enzyme for viral replication, and tankyrase 2 (TNKS2), a key regulator in Wnt (a portmanteau of ``Wg'' and ``int'' standing for ``Wingless-related integration site'') signalling and a major cancer drug target. Utilizing datasets of 16,780 and 21,702 compounds\cite{Loeffler2024} respectively, we assess the model’s structural fidelity via root mean square deviation (RMSD) and local distance difference test (LDDT) comparisons against traditional docking, while simultaneously benchmarking its affinity predictions against binding free energies derived from the ESMACS protocol. The ensemble strategy of ESMACS effectively mitigates the sensitivity of single-trajectory simulations to initial conditions and enables UQ assessment which is entirely absent in standard AI methods. It therefore provides a robust benchmark for assessing the performance of the AI-driven affinity predictors. By examining both global correlations and the precision of top-ranked candidates, we aim to establish whether Boltz-2 can serve as a reliable, high-speed surrogate for physics-based pipelines in virtual screening phase of drug discovery.

\section{Results}\label{sec2}

In this section, we present a quantitative assessment of Boltz-2’s performance across two distinct protein targets, 3CLPro and TNKS2. Our evaluation focuses on two critical dimensions of the model's output: the geometric fidelity of the predicted protein-ligand complexes and the accuracy of the estimated binding free energies. To establish a rigorous benchmark, we first perform a global analysis across the entire dataset of 38,482 compounds, comparing Boltz-2’s structural poses with traditional docking results and its affinity predictions with ESMACS energies. This is followed by a targeted analysis of the top 100 compounds identified by Boltz-2, allowing us to determine the model's reliability in high-confidence lead identification and its potential as a high-throughput alternative to computationally expensive physics-based methods.

\subsection{Global Evaluation of Structure and Binding Affinity}

We first assess the generalizability of Boltz-2 across the full datasets of 16,780 (3CLPro) and 21,702 (TNKS2) compounds. We compared the 3D complex structures predicted by Boltz-2 with traditional docking poses. Structural similarity was quantified using the root mean square deviation (RMSD). The affinity scores from Boltz-2 were compared with the ESMACS binding free energies (\(\Delta G\)).

\subsubsection{Structural comparison}

To evaluate Boltz-2's structural predictions and their reliability, we first examine the predicted protein conformations and ligand binding poses, comparing them with x-ray structures and conventional docking results. We then illustrate representative binding sites to highlight system-specific variations and potential alternative binding modes. Finally, we assess the internal consistency of Boltz-2 predictions using the confidence score, which provides a quantitative measure of pose reliability across the ligands. 

\begin{figure}[H]
  \centering
  \includegraphics[width=1.0\textwidth]{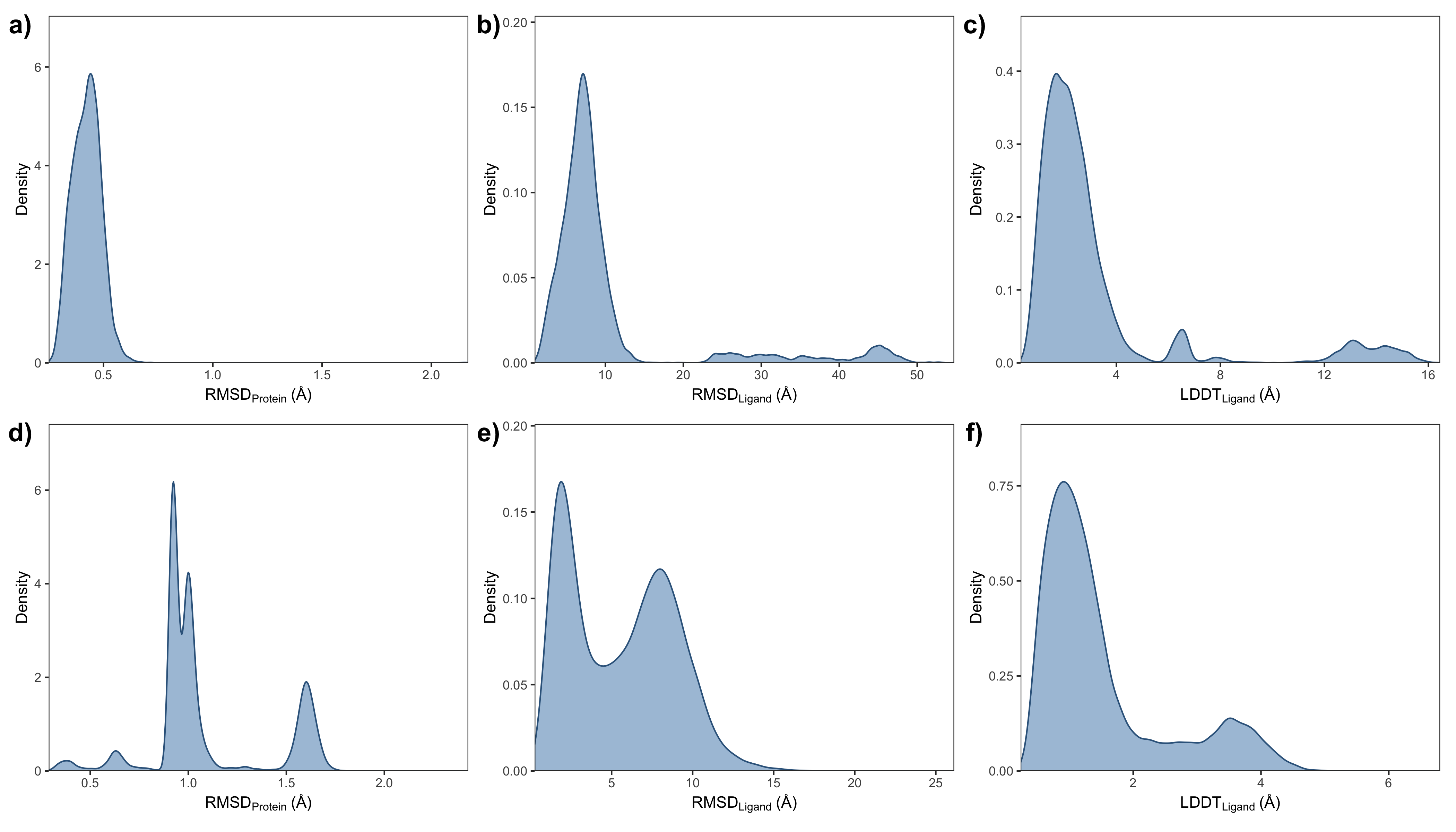}
  \caption{Structural comparison of Boltz-2 and docking predictions for 3CLPro (a-c) and TNKS2 (d-f). Distributions are shown for pairwise RMSDs of the protein structures (a, d) and ligand poses (b, e), alongside LDDT scores (c, f) evaluating the preservation of the local atomic environment and binding site interactions.}
  \label{fig:rmsd_lddt}
\end{figure}

\paragraph{Protein structure comparison}
We compare the protein structures predicted from Boltz-2 with the x-ray structures, and the ligand binding poses from Boltz-2 and the docking approach. The comparative distributions are presented in Fig. \ref{fig:rmsd_lddt}. As a single x-ray structure (6W63 for 3CLPro and 4UI5 for TNKS2) is used in the docking approach, the Root-Mean-Square Deviations (RMSDs) for the protein reflect the variations of the predicted protein structures. A single protein conformation is predicted for 3CLPro from Boltz-2 prediction, which closely resembles the x-ray structure (RMSD $\approx 0.4$~\r{A}). For TNKS2, substantial variations are observed in Boltz-2 predicted protein conformations, as evidenced by multiple peaks in the RMSD distribution. While a small portion of the predicted structures assemble the x-ray structure, the dominant conformations are centred at 1.0 \r{A} with some variations that extend RMSD to ${\sim}1.8$~\r{A} from the x-ray structure.

\begin{figure}[H]
  \centering
  \includegraphics[width=0.7\textwidth]{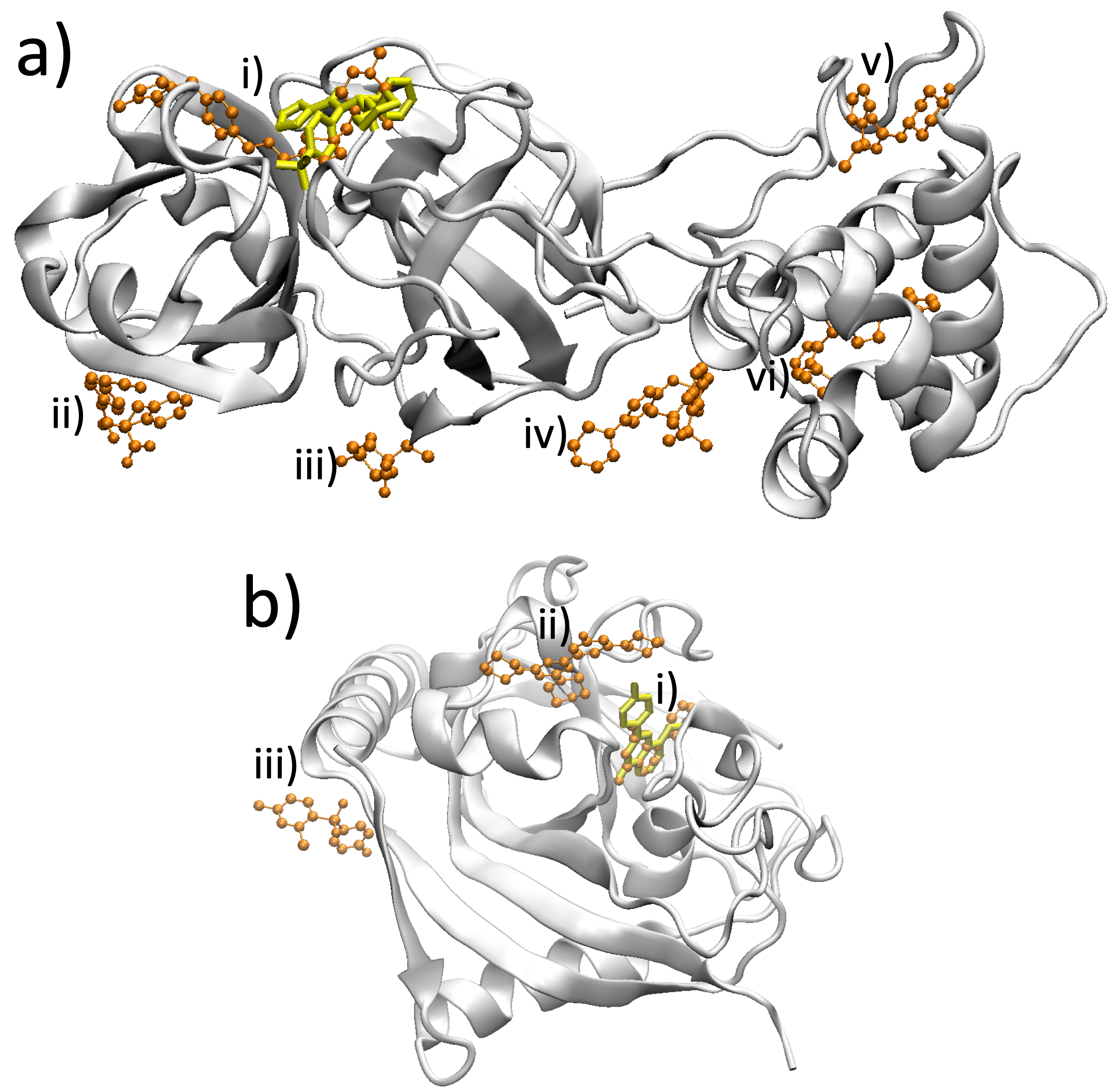}
  \caption{Representative binding sites predicted by Boltz-2 co-folding for a) 3CLPro and b) TNKS2. The protein (cartoon) and a bound compound (yellow sticks) are from PDB 6W63 and 4UI5, respectively. Five binding sites (i-vi) are predicted for 3CLPro, with LDDT metric values of 1.8 \r{A}, 6.4 \r{A}, 7.8 \r{A}, 13.0 \r{A}, 13.2 \r{A} and 14.3 \r{A} (Fig. \ref{fig:rmsd_lddt}c), respectively. For TNKS2, all but one compounds bind at or near the x-ray identified binding site (i, ii), with the exception at a different site (iii), with LDDT values (Fig. \ref{fig:rmsd_lddt}f) of 0.8 \r{A}, 3.6 \r{A} and 6.9 \r{A}, respectively.}
  \label{fig:poses}
\end{figure}

The RMSD for the ligands show differences in their positions when complexed with the protein. As the binding site is explicitly defined in the docking approach, the ligands will always be in the binding site as defined in the x-ray structure. The predicted locations from Boltz-2 for some ligands, however, may not be in or near the binding sites as defined in the x-ray structure (see Fig. \ref{fig:poses}), resulting in large RMSDs. It should be noted that, when spatial constraints were applied, the probability of compounds being found in the catalytic pocket region can be significantly increased\cite{Boltz2_constraints}.

\paragraph{Ligand binding pose comparison}
The analysis of ligand binding poses reveals significant system-dependent variability. For 3CLPro, the RMSD values of the ligand span a wide range, with a primary peak below 10 \r{A} and a plateau between 22 \r{A} and 50 \r{A}. In contrast, TNKS2 ligands show a dual-mode distribution, with both peaks residing below 10 \r{A}, indicating a higher accuracy of Boltz-2 predicted structures for TNKS2 than for 3CLPro when compared to docking-derived poses. The structural accuracy of predicted protein-ligand complexes is further assessed by the local distance difference test (LDDT). Unlike global RMSD metrics, LDDT is superposition-free which makes it more robust when evaluating local binding site details. The LDDT metrics mirror the shape of the ligand RMSD distributions but with metric values roughly third. LDDT is less affected by the global shifts of the ligands, and will give a high score (small LDDT values) if the ligand preserves its internal geometry and its interactions with the binding pocket, whereas ligand RMSDs might be penalise heavily (large ligand RMSDs).
For TNKS2, most of the LDDT metric values fall below 5 \r{A} (see Fig. \ref{fig:rmsd_lddt}f), and the second peak is significantly reduced from that in the ligand RMSDs (Fig. \ref{fig:rmsd_lddt}e), indicating general agreement on the approximate location of the binding site. For 3CLPro, about 11.3\% of the ligands demonstrate LDDT metrics larger than 6 \r{A} (Fig. \ref{fig:rmsd_lddt}c), indicating that the co-folding approach of Boltz-2 predicts binding sites that diverge significantly from the experimentally determined native site (Fig. \ref{fig:poses}).

The superior pose agreement for TNKS2 is likely attributable to its narrower and more well-defined binding pocket compared to the more flexible and ambiguous binding site of 3CLPro. The observed discrepancies, particularly the second peak for TNKS2 and the wide plateau for 3CLPro in the ligand RMSD plots, arise from distinct phenomena for each protein. For 3CLPro, the similar shape of the ligand RMSD and LDDT metrics distributions suggests that the peaks and the plateaux with high values correspond to ligands predicted to bind in entirely different sites. For TNKS2, the majority of the compounds bind within or adjacent to the site identified in the x-ray structure, while one exception interacts with an alternative site (Fig. \ref{fig:poses}b). The significant reduction in the second peak amplitude from the protein-centric proximity profiles (LDDT in Fig. \ref{fig:rmsd_lddt}f) indicates that some high values of ligand RMSDs are dominated by incorrect orientations, such as rotated or flipped poses, within the correct binding pocket. This distinction underscores a critical limitation in Boltz-2’s ability to resolve protein-ligand interactions precisely, as the model successfully identifies the binding site but fails to accurately orient the ligand within the local topology.

\paragraph{Reliability estimation in binding pose from Boltz-2}

To further characterise the reliability of these predicted ligand poses, we next examine Boltz-2's internal confidence scores, which quantify the model’s self-assessed consistency across the ligands. The confidence score is a normalized value between 0 and 1, with higher scores indicating greater consistency of the predicted interaction according to the model’s learned representations. Importantly, this score does not correspond to experimental binding affinity, but serves as a relative metric for filtering and prioritising the algorithm's own predictions.

\begin{figure}[H]
  \centering
  \includegraphics[width=0.9\textwidth]{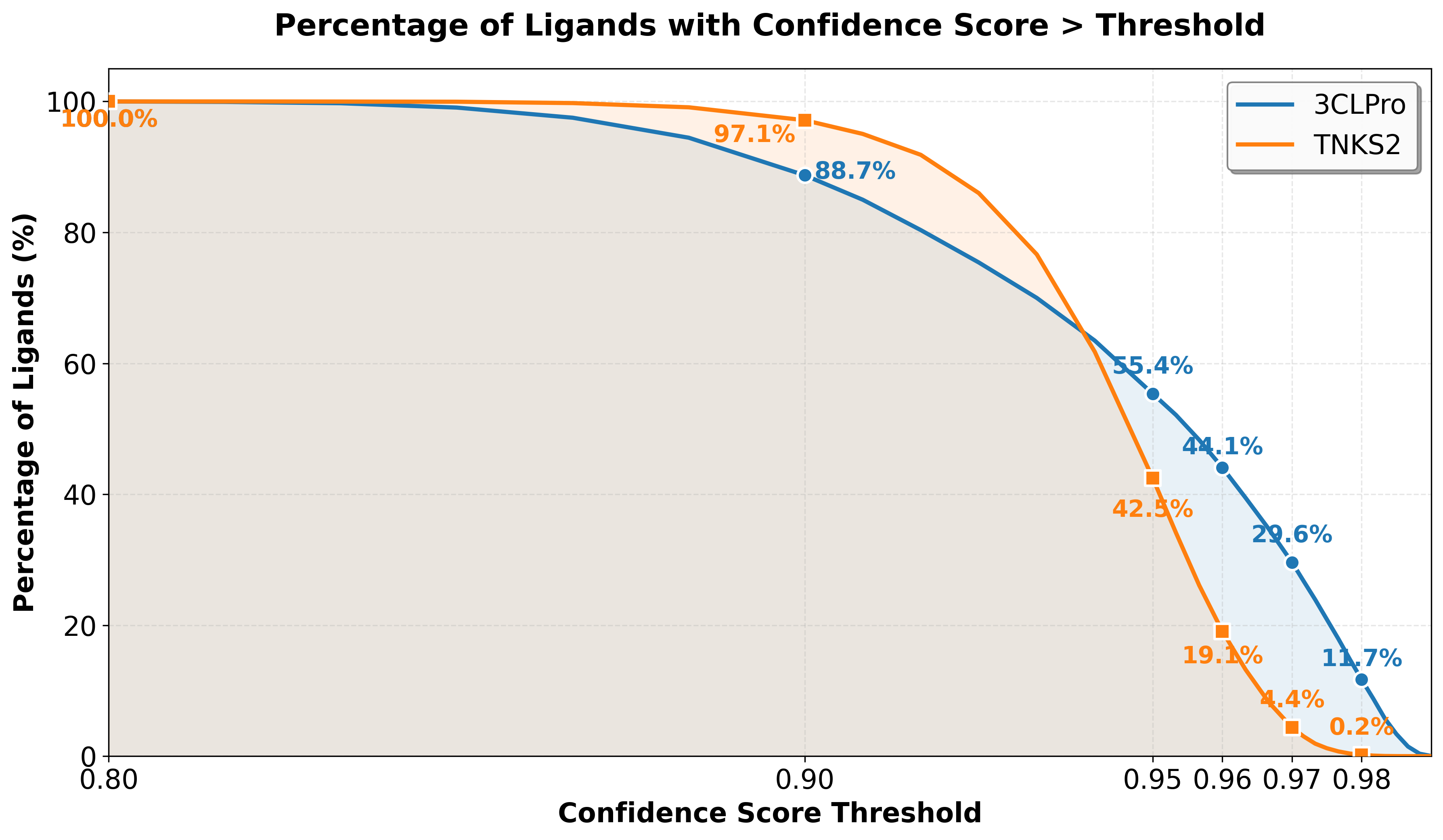}
      \caption{The distribution of confidence scores from the Boltz-2 predictions of both 3CLPro and TNKS2 systems. Percentages are calculated relative to the total number of ligands in each protein system.}
  \label{fig:confidence score of Boltz2}
\end{figure}

To characterise predictive uncertainty, we analysed the distribution of confidence scores and quantified the proportion of ligands exceeding a series of increasingly stringent thresholds. For each system, we calculated distribution statistics and the cumulative proportion of ligands with confidence scores above specified thresholds, with results shown in Fig. \ref{fig:confidence score of Boltz2}. We observed that Boltz-2 confidence scores are strongly compressed for both systems. All ligands scored above 0.8, indicating that low-confidence predictions are essentially absent in this dataset and that thresholds below 0.9 provide little discriminatory power. The distributions are highly concentrated around similar means (0.946 for 3CLPro and 0.944 for TNKS2), suggesting that Boltz-2 assigns uniformly high internal confidence across large ligand libraries. In other words, as with all ML/AI methods, it makes overconfident predictions.

Despite this compression, additional structure emerges at more stringent thresholds. For 3CLPro, 55.35\% of ligands exceed a confidence score of 0.95, decreasing to 44.08\% above 0.96 and 29.61\% above 0.97. In contrast, TNKS2 exhibits a markedly sharper drop, with only 42.46\%, 19.07\%, and 4.38\% of ligands remaining above the same respective thresholds. Thus, although TNKS2 contains a larger ligand set overall, Boltz-2 assigns high-confidence predictions to a substantially smaller fraction of compounds once strict filtering is applied.

These results reveal target-dependent behaviour in Boltz-2's internal uncertainty estimates. Although the confidence score is compressed at coarse levels, subdivision within the high-confidence regime ($\ge 0.95$) exposes meaningful separation between ligands and between targets. 

Overall, our analysis reveals notable divergences between the structures predicted by Boltz-2 and docking methods (see Fig. \ref{fig:rmsd_lddt}). This is particularly evident for 3CLPro, where a non-negligible percentage of ligands are predicted to bind at a site different to the one observed from x-ray crystallography. We cannot exclude the possibility that certain compounds, though screened or designed for the primary binding site, may occupy alternative binding sites. The more favourable binding energies observed for docked poses in our focused study (see Section \ref{sec:energies_top100}) suggest that docking-derived structures may be more reliable. Ultimately, such AI-generated structures must be rigorously validated through experimental structural determination\cite{terwilliger_alphafold_2024}. Indeed, a previous report has identified cases where Boltz-2 predicted PI3K-\(\alpha\) inhibitors to bind in regions entirely distal to their confirmed orthosteric site \cite{Boltz2_blog_ryan}, further underscoring the inherent risks of structural misidentification in co-folding models. This indicates yet again that caution is warranted when interpreting structures from AI-based methods. In parallel, the Boltz-2 confidence scores provide an internal measure of reliability across ligands. For both 3CLPro and TNKS2, the scores are generally high: all ligands exceed 0.80, and approximately 90\% surpass 0.90. However, unless thresholds are set stringently above 0.95, the confidence score shows limited discriminatory power within these datasets. This highlights the range within which the metric is informative for assessing prediction reliability for the two systems.

\subsubsection{Boltz-2 predicted binding affinities}

While models such as AlphaFold 3 focus primarily on structural coordinates, Boltz-2 stands out through its claimed ability to predict binding affinities with precisions that approach rigorous physics-based methods. Recent benchmarks\cite{Boltz2_blog_ryan} suggest it excels at qualitative ranking, effectively differentiating between potent and weak analogues within a chemical series during lead optimisation. Here, we evaluate these capabilities by assessing the linear relationships and ranking fidelity of Boltz-2 relative to physics-based binding free energy calculations, utilizing Pearson correlation (\(r\)) and Spearman rank correlation (\(\rho \)), respectively.

Our previous studies across diverse molecular systems have demonstrated that ESMACS reliably ranks binding affinities, despite a characteristic overestimation of absolute free energies. While no universal threshold exists due to the system-dependent nature of the overestimation, compounds with ESMACS values exceeding \(-20\)~kcal/mol are generally considered non-binders or extremely weak hits\cite{MMPBSA_overestimate}. In contrast, Boltz-2 exhibits a ``regression to the centre'' effect, where predictions are concentrated within a remarkably narrow range, with the vast majority of affinities falling between \(-5\) and \(-8\)~kcal/mol. The model maintains these relatively strong predicted affinities even for probable non-binders, indicating an evident lack of sensitivity in distinguishing true hits from decoys (non-binding molecules sharing similar 1D or 2D characteristics with the hits) based solely on the predicted binding affinities. Consequently, both \(\Delta G_{\text{Boltz}}\) and \(\Delta G_{\text{ESMACS}}\) are subject to inherent, disparate errors and exhibit significant numerical misalignment, making their absolute values not directly comparable. To account for this, we employed standardised major axis (SMA) regression for linear fitting. Unlike ordinary least squares (OLS), which assumes error only in the dependent variable and often suffers from slope attenuation (regression dilution), SMA minimises the area of right triangles formed by both vertical and horizontal residuals. This ensures the fitted line remains centrally aligned within the data cloud, providing a more symmetric representation of the relationship between two error-prone variables.

\paragraph{Correlation of predicted binding affinity between ESMACS and Boltz-2}

\begin{figure}[H]
  \centering
  \includegraphics[width=1.0\textwidth]{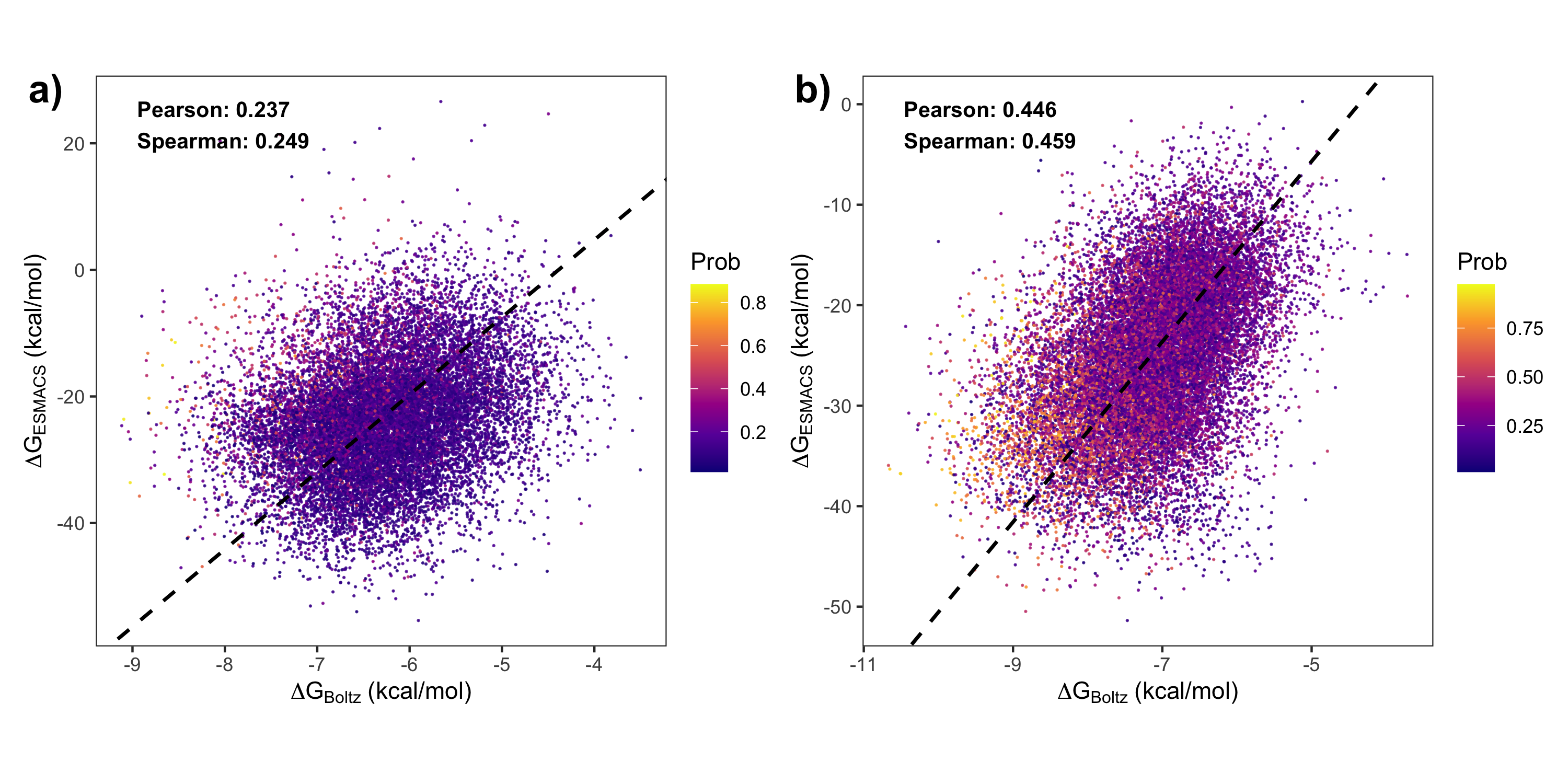}
  \caption{Correlation between binding free energies predicted by Boltz-2 (\(\Delta G_{\text{Boltz}}\)) and calculated via ESMACS (\(\Delta G_{\text{ESMACS}}\)) for a) 3CLPro and b) TNKS2. Dashed lines indicate standardised major axis (SMA) regressions. Data points are coloured by Boltz-2 predicted binding probabilities.}
  \label{fig:energy_Boltz_ESMACS}
\end{figure}

While Boltz-2 yields a moderate correlation for TNKS2 (Pearson \(r=0.45\) and Spearman \(\rho=0.46\), see Figure \ref{fig:energy_Boltz_ESMACS}), it exhibits more limited accuracy for 3CLPro (\(r=0.24\) and \(\rho=0.25\)). This discrepancy can be attributed to distinct binding site architectures: while 3CLpro possesses a large and branched binding site that introduces conformational and energetic variability in ligand bindings, TNKS2 features a narrow and well-defined binding pocket that enables more reliable pose and energy determination. The structural differences in the binding sites may also account for the more accurate prediction of binding probabilities for TNKS2. As shown in Fig. \ref{fig:energy_Boltz_ESMACS}b, TNKS2 exhibits a visually clearer clustering of high-probability points within the high-affinity (more negative \(\Delta G\)) region compared to 3CLPro. However, only weak correlations are observed between the Boltz-2 predicted binding affinities and binding probabilities, with Pearson correlation coefficients of 0.21 and 0.25 for 3CLPro and TNKS2, respectively. Almost all compounds are predicted by Boltz-2 to have binding affinities below $-4$~kcal/mol, even though many are assigned very low binding probabilities. Despite their different roles as classification and regression heads, the binding affinity and binding probability derive from the same structural embeddings and should manifest a consistent correlation between their outputs. The weak correlation observed in the current study indicates an inconsistency in the model's multi-objective architecture.

\paragraph{Reproducibility of Boltz-2}

To evaluate internal consistency, Boltz-2 predictions were compared between independent inference runs for each protein system. We used the same default pre-trained model weights for these runs, and hence the aleatoric uncertainties from random seeds during its trainings were not accounted for in our repeated runs.

\begin{figure}[H]
  \centering
  \includegraphics[width=1.0\textwidth]{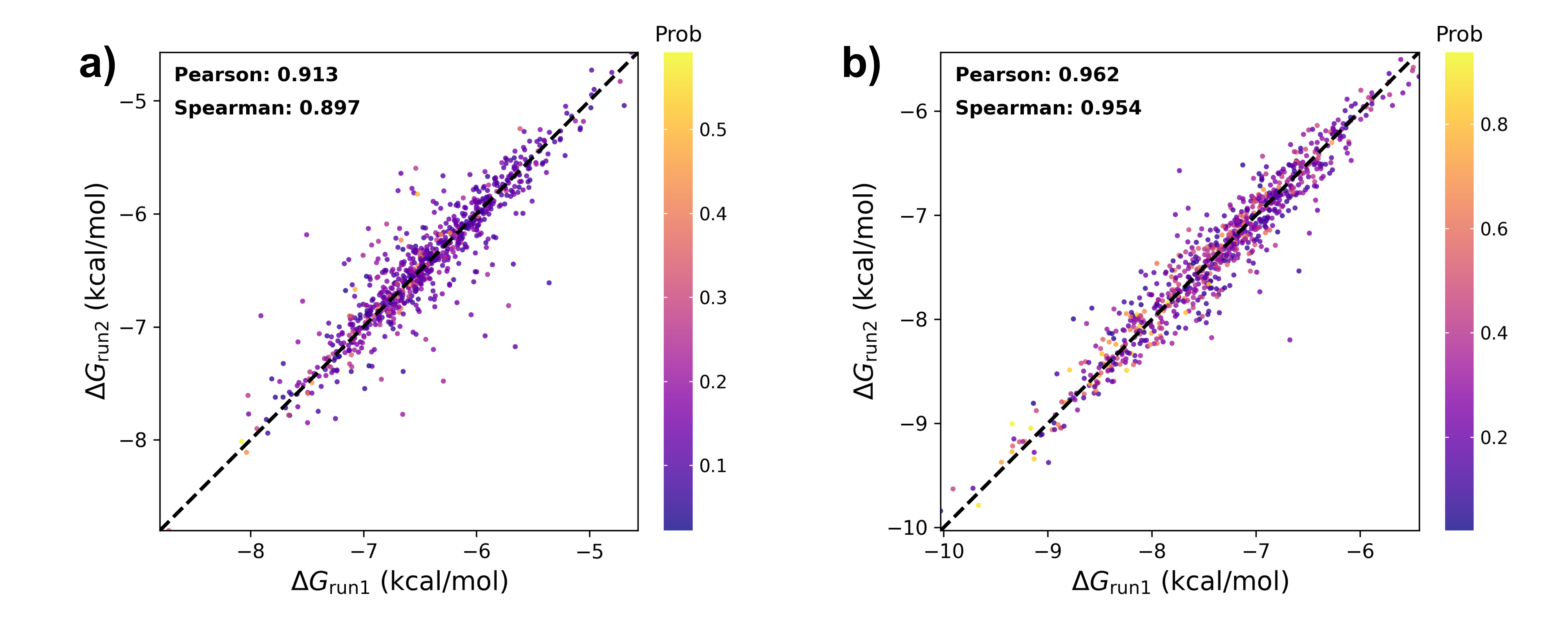}
  \caption{Binding affinity consistency between two independent runs of Boltz-2 for a) 3CLPro and b) TNKS2.}
  \label{fig:Binding_affinity_consistency}
\end{figure}

For this study, 793 ligands for 3CLPro and 799 ligands for TNKS2 were independently re-evaluated with Boltz-2. Reproducibility was quantified using Pearson and Spearman correlation between the two runs. As shown in Fig. \ref{fig:Binding_affinity_consistency}, both 3CLPro and TNKS2 predictions showed high reproducibility, with Pearson \(r\) of 0.913 and 0.962, and Spearman $\rho$ of 0.897 and 0.954, respectively, indicating that Boltz-2 predictions are generally stable. Meanwhile, a slight system-dependent difference was observed, with TNKS2 showing better alignment between runs than 3CLPro. Nevertheless, despite the strong overall correlation, absolute differences in predicted binding affinities between runs can reach ${\sim}1.5$ kcal/mol. Such deviations are significant for compound ranking, potentially shifting the priority of lead candidates within the model's relatively narrow prediction range (approximately $-4$ to $-9$ kcal/mol for the majority of ligands).

\subsection{Precision Analysis of Top-100 Boltz-2 Predictions}

To evaluate the utility of Boltz-2 as a primary filter for virtual screening, we analysed the top 100 compounds ranked according to the predicted affinity of Boltz-2 for each target. It should be noted that the overlap between the top 100 compounds identified by Boltz-2 and ESMACS was minimal, with only 2 and 1 compound(s) in common for 3CLPro and TNKS2, respectively. Statistical analysis reveals that for datasets of these sizes (\(N=16,780\) and \(21,702\)), the probability of finding at least one common compound by random chance is \(45.0\%\) and \(36.9\%\), respectively. This suggests that the two scoring methodologies explore fundamentally different regions of the chemical landscape.

\subsubsection{Binding poses from Boltz-2} \label{sec:saturation}
During the evaluation of Boltz-2 generated poses, a systematic discrepancy was observed between the protonation and saturation states of the predicted ligand structures and their canonical biochemical forms. The Boltz-2 predicted structures contain only heavy atoms (C, N, O, S, etc.). Explicit hydrogen atoms were added with OpenEye tools to these structures. For a subset of these top 100 compounds, 28 for 3CLPro and 30 for TNKS2, the atomic models exhibited divergent hydrogen counts and corresponding indices of hydrogen deficiency (IHD) relative to their defining SMILES strings.

\begin{figure}[H]
  \centering
  \includegraphics[width=0.7\textwidth]{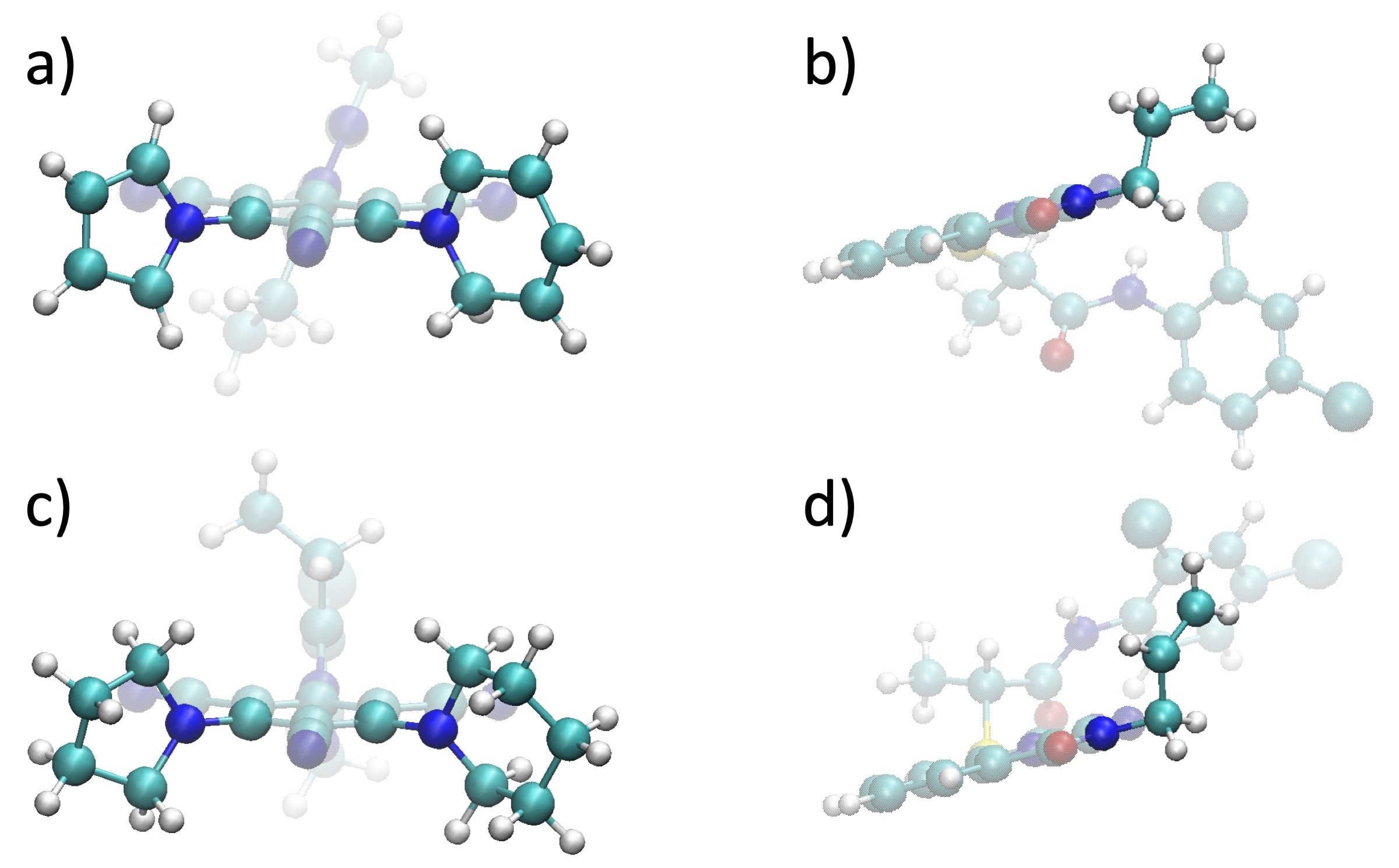}
  \caption{Comparison of saturation state deviations. Boltz-2 saturation is derived from hydrogen placement on predicted heavy-atom coordinates. Representative deviations of the saturation states from Boltz-2 predictions are shown for rings (a) and aliphatic chains (b), alongside their SMILES string counterparts (c, d).}
  \label{fig:saturation}
\end{figure}

This deviation manifests primarily in two distinct patterns: 1) for ring systems, predicted structures frequently show increased unsaturation. Aromatic or alicyclic rings in the SMILES are often predicted in a more unsaturated, often fully conjugated, state. For example, a pyrrolidinyl moiety (saturated, IHD=1) is predicted as a pyrrolyl group (aromatic, IHD=3) and a piperidinyl moiety (saturated, IHD=1) as a dihydropyridinyl group (IHD=3) (see Fig. \ref{fig:saturation}a,c); and 2) for aliphatic chains, Boltz-2 predicted models display a bias toward increased saturation. Unsaturated carbons (e.g., vinyl group with IHD=1, Fig. \ref{fig:saturation}d) are sometimes predicted as saturated $sp^3$-hybridized ones (e.g., ethyl group with IHD=0, Fig. \ref{fig:saturation}b), effectively adding hydrogen atoms not present in the original specification.

We hypothesise that these opposing biases stem from Boltz-2's training on structural datasets and their implicit energy landscape. For rings, the model may favour planar, conjugated geometries that represent deep local minima in vacuum or implicit-solvent potentials. For chains, the model appears to prioritize sterically relaxed, tetrahedral geometries, potentially smoothing out conjugated patterns that are stabilized specifically by protein-ligand interactions or explicit electronic effects. Passaro \textit{et al.}\cite{Boltz2_2025} explicitly list some significant physical issues with the poses Boltz-2 generates, including ``slightly incorrect bond lengths and angles, incorrect stereochemistry at chiral centers and stereobonds and aromatic rings predicted to be non-planar''. This is precisely the reason for the saturation state deviations we observe here.

These systematic alterations in ligand bonding topology have direct consequences for binding affinity predictions. Changes in hybridization and saturation often substantially affect molecular shape, steric complementarity with the binding pocket, electrostatic potential, hydrogen-bonding capacity, and local flexibility and entropic contributions to binding, all critical determinants of binding free energy. To isolate the source of prediction error, we performed two parallel sets of affinity calculations: 1) Using the Boltz-2 predicted structures with hydrogens added according to the model's implicit bonding topology; and 2) Using canonical and re-protonated structures, where the heavy-atom coordinates from Boltz-2 were retained, but all bonding and hydrogenation states were corrected to match the definitive SMILES representation. This dual-pathway approach allows us to distinguish between errors arising solely from 3D pose geometry and those compounded by incorrect chemical identity. The comparative analysis of these two affinity sets, presented in Section \ref{sec:energies_top100}, quantifies the impact of saturation state artifacts on the overall performance of the Boltz-2 pipeline.

\subsubsection{Binding free energies}\label{sec:energies_top100}

\begin{figure}[H]
  \centering
  \includegraphics[width=1.0\textwidth]{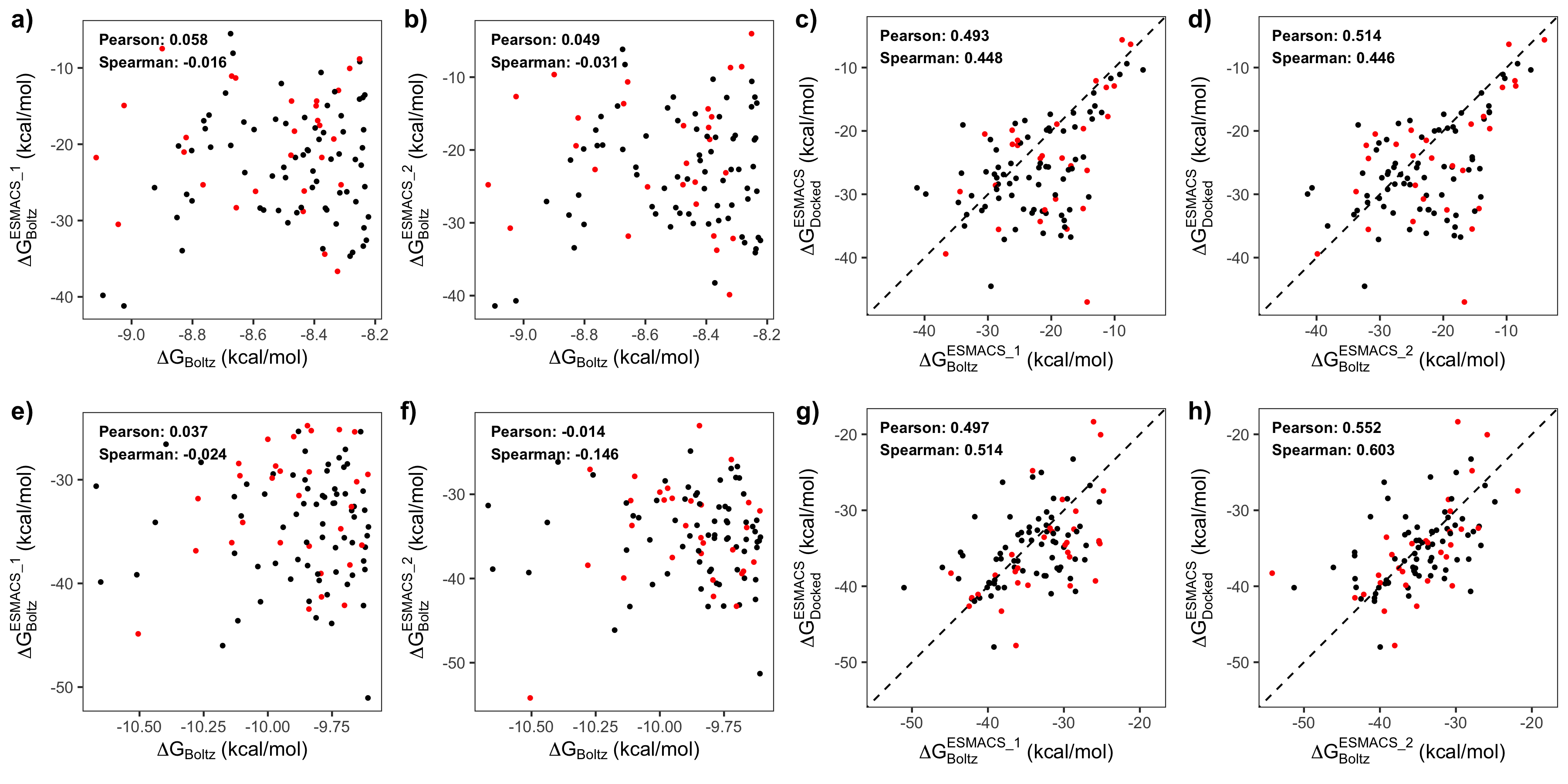}
  \caption{Comparison of binding free energies from Boltz-2 prediction ($\Delta G_{Boltz}$) and ESMACS simulations. Three sets of ESMACS simulations are performed, denoted as $\Delta G_{Boltz}^{ESMACS\_1}$, $\Delta G_{Boltz}^{ESMACS\_2}$ and $\Delta G_{Docked}^{ESMACS}$, respectively, initiated from structures derived directly from Boltz-2 predicted structures, derived from SMILES strings with positions of heavy atoms from Boltz-2 predictions, and from docked approach. Some compounds have different saturation or protonation states for $\Delta G_{Boltz}^{ESMACS\_1}$ and $\Delta G_{Boltz}^{ESMACS\_2}$ calculations; they are shown as red dots.}
  \label{fig:dG_top100}
\end{figure}

Although there are weak to moderate overall correlations for the entire datasets for 3CLPro and TNKS2 (Fig. \ref{fig:energy_Boltz_ESMACS}), no correlation is observed between the initial Boltz-2 predictions and the ESMACS calculations for the top 100 compounds selected (Fig. \ref{fig:dG_top100}a,b,e,f), no matter what saturation (Fig. \ref{fig:saturation}) and/or protonation states assigned for the compounds. The Pearson and Spearman correlation coefficients are close to zero. It is not uncommon for correlations to ``break down'' when zooming in on the top-performing candidates. The selection of top 100 compounds significantly reduces the variance of the Boltz-2 predicted binding affinities (within 1.0 kcal/mol of each other), which is likely to be smaller than the model's inherent uncertainty, making the correlation mathematically collapses. Furthermore, the reported insensitivity of Boltz-2 affinities to structural accuracy\cite{Bret_2026} explains the lack of correlation with our ESMACS calculations which responds to the specific physical environment of the binding site. Similar to other co-folding approaches that do not explicitly account for physics-based interactions during structure prediction\cite{masters_co_folding_2025}, Boltz-2 appears to derive binding affinities from features largely independent of the final ligand pose\cite{Bret_2026}. Therefore, the binding affinities predicted from Boltz-2 are not a reliable quantitative predictor of the more rigorous, physics-based ESMACS binding free energies for this specific set of 100 compounds. This indicates that Boltz-2 may be effective for classification (identifying binders), but does not provide accurate binding free energies compared to simulation-based approaches.

There is a moderate positive correlation between two different ESMACS calculations that use different initial structural inputs (Boltz-generated vs. docked structures), with correlation coefficients ranging from approximately 0.45 to 0.60 (Fig. \ref{fig:dG_top100}c,d,g,h). Therefore, the ESMACS calculations are relatively consistent, showing moderate agreement between runs initiated with different, but likely viable, binding poses. The moderate correlation suggests general agreement in binding strength, although specific values may vary slightly. Compounds with different saturation and/or protonation states do not appear systematically as major outliers, although some deviation is present (Fig. \ref{fig:dG_top100}c,d,g,h). The corrected saturation or protonation states improve the correlations for TNKS2, although their impact is less significant for 3CLPro.

While the data points cluster reasonably well around the diagonal dashed line (line of identity, where x=y), there are more data points below the line: 66 out of 100 compounds for 3CLPro (Fig. \ref{fig:dG_top100}c) and 60 out of 100 for TNKS2 (Fig. \ref{fig:dG_top100}g). The correction for saturation and protonation scarcely changes the trend, with 67 and 59 data points below the line for 3CLPro and TNKS2 (Fig. \ref{fig:dG_top100}d,h), respectively. It is important to note that these compounds were originally identified through an iterative active learning protocol for optimal molecular design\cite{Loeffler2024}, which coupled the generative AI framework REINVENT with the physics-based ESMACS approach. Both REINVENT and ESMACS serve as core components within the IMPECCABLE workflow, and the current results indicate that for these expertly designed molecules, the docked structures consistently yield more favourable binding free energies than those predicted by Boltz-2. 

\section{Discussion}\label{sec3}

We have conducted a rigorous evaluation of the Boltz-2 foundation model's capability to predict protein-ligand structures and binding affinities across two large-scale datasets, 3CLPro and TNKS2. By benchmarking the model's performance against traditional docking and the physics-based ESMACS protocol, we have identified several critical limitations within the current state of AI-driven structural biology, which are summarised here.

Our structural analysis reveales significant global RMSD variations in both protein and ligand conformations. These findings indicate that, while Boltz-2 captures a broader conformational ensemble than traditional static docking, it does not consistently identify a single, dominant ``native'' binding site across diverse chemical libraries.

Regarding binding energetics, Boltz-2 demonstrates only weak to moderate correlations with physics-based free energy values in the global evaluation. More strikingly, the focused analysis of the top 100 compounds revealed a complete lack of correlation with fine-grained ESMACS results. This indicates that, while Boltz-2 is efficient at broad-spectrum screening, it lacks the resolution required for the hit-to-lead phase, where ranking structural analogues to establish clear structure-activity relationship is critical. The observation of hydrocarbon ``saturation'' issues for certain ligands further indicates that the model is not able to accurately predict the binding poses for some ligands, and consequently their binding affinities which heavily rely on the structures it predicts.

The root of these limitations lies in a fundamental assumption most machine leaning architectures rely on: that biomolecular data behaves as a smooth, differentiable manifold. Rather than explicitly incorporating the underlying physics of biomolecular interactions, ML-based approaches rely on statistical patterns from training datasets. The assumption of differentiability often fails when navigating the vast chemical space, where structure-activity relationships (SARs) are often discontinuous. A primary manifestation of this is the ``activity cliff''\cite{activity_cliff}, where minor structural modifications lead to disproportionate changes in binding affinity. Such non-linear phenomena are inherently difficult for current AI architectures to navigate, underscoring the necessity of integrating fundamental physics to both interpret model failures and develop more robust algorithms\cite{AI_needs_physics_coveney2026}.

In conclusion, while Boltz-2 offers an efficient tool for generating protein–ligand complexes when canonical docking approaches are unfeasible, its poor performance in affinity ranking indicates that it cannot serve as a reliable substitute for high precision, physics-based methods in late-stage drug discovery. Our results suggest a fundamental incompatibility between the model's smooth, differentiable assumptions and the non-linear, discontinuous nature of the physical energy landscape. Rather than learning the underlying physics of molecular recognition, the model appears to rely on \textit{ad hoc} statistical heuristics that are largely decoupled from local atomic environments. This vulnerability is exacerbated by the ``deluge of spurious correlations'' inherent in massive datasets. As demonstrated by Calude and Longo\cite{calude_deluge_2017}, the ratio of false to true correlations increases combinatorially as a result purely of the size of a dataset, not its nature. One needs to find ways of identifying the true correlations to weed out all the false ones. This requires scientific understanding applied by a form of reinforcement learning. Until foundation models can move beyond these mathematical artifacts to incorporate the rigour and uncertainty quantification of physics-based interactions, they will remain insufficient for navigating the complex activity cliffs of real-world chemical space.

\section{Methods}\label{sec4}

We describe in this section the datasets of compounds used for the benchmarking of the Boltz-2 foundation model against established physics-based methodologies, the execution of the ensemble-based ESMACS protocol for affinity free energy calculations and the specific metrics employed to quantify the structural and energetic divergence between AI-driven and traditional approaches.

\subsection{Dataset preparation}

The datasets for 3CLPro and TNKS2 were obtained from a large-scale compound library previously described by Loeffler \textit{et al}\cite{Loeffler2024}. While full details are available in the original study, the generation process is summarized briefly below.

The two datasets consist of initial $\sim$10,000 compounds each, screened from ZINC15 and MCULE compound databases using a surrogate docking model\cite{Clyde_2023} for 3CLPro, or generated using REINVENT\cite{reinvent_2024}, a generative molecular AI, based on 27 known compounds with available experimental affinity measurements for TNKS2. More compounds were then generated iteratively with REINVENT, which uses reinforcement learning (RL), to optimise molecules subjected to external ``information'' i.e., scoring functions which evaluate each compound for its fitness. The scoring function used was the predicted binding affinities from ESMACS simulations (see subsection \ref{sec:esmacs}). In each generative active learning (GAL) iteration, a new batch of compounds was produced with improved fitness. Using various batch sizes in each GAL cycle and the number of cycles to reach convergent behavior toward the end of GAL, approximately 16,900 and 21,861 compounds were studied in total for 3CLPro and TNKS2, respectively. We removed the compounds that encountered execution errors during the molecular dynamics simulation or the free energy calculation steps, resulting in final datasets of 16,780 and 21,702 compounds for the two molecular systems ($\sim$0.7\% failures).

\subsection{Boltz-2 Inference and Structure Prediction}

Predictive modelling and binding affinity prediction were conducted using Boltz-2. For each ligand in the 3CLPro and TNKS2 datasets, the model was provided with the protein sequence and SMILES strings for the ligands. Unlike traditional docking, Boltz-2 utilizes a co-folding architecture that allows for reciprocal conformational changes in both the protein and ligand during the prediction process.

Inference was performed using the default pre-trained model weights, with no spatial constraints applied, to evaluate its ``out-of-the-box'' performance on large-scale screening tasks. For each compound, we extracted the predicted binding affinity (reported as a continuous score) and the 3D atomic coordinates of the pose with the highest confidence. Predicted binding affinities, reported as $\!\log_{10}(\text{IC}_{50})$ values with $\text{IC}_{50}$ in micromolar units (\(\mu\)M), can be converted to pIC50 in \(\text{kcal/mol}\) using formula $(6 - \!\log_{10}(\text{IC}_{50})) * 1.364$. 

Boltz-2 predictions were performed on Isambard-AI which is made up of 5,448 NVIDIA GH200 Grace-Hopper superchips at the University of Bristol. In order to improve computational efficiency, UniProt Reference Clusters at 90\% sequence identity (UniRef90) \cite{uniprot2025uniprot} was used to generate local multiple sequence alignments. UniRef90 is a clustered protein database from UniProt, where sequences with no less than 90\% identity are merged to reduce redundancy while retaining representative diversity for the known protein sequence space. Boltz-2 inference required approximately 77 seconds per compound for 3CLPro and 74 seconds for TNKS2 to predict protein-ligand structures and binding affinities on a single GH200 GPU. To accelerate the screening process, we used 32 GPUs concurrently, providing a proportionate reduction in the total wall-clock time for the full dataset. 

\subsection{ESMACS calculations} \label{sec:esmacs}

The initial ESMACS binding free energies were extracted from our previous molecular design study\cite{Loeffler2024}, in which a coarse-grained ESMACS protocol was employed to optimise computational efficiency. The standard ESMACS protocol\cite{ESMACS_2015} uses an ensemble of 25 replicas and 4 ns production run for each compound-protein complex to get precise predictions. For a drug screening project, like our previous study\cite{Loeffler2024}, tens of thousands of compounds need to be evaluated. A coarse-grained protocol (CG-ESMACS), using an ensemble of 10 replicas and smaller buffer distance of 10 \r{A}, is deployed, which inevitably decreases the precision of the predictions but is well-suited for high-throughput virtual screening. Note that the time-consuming entropic calculation is omitted in the current ESMACS protocol as its inclusion, although brings the calculated binding free energies closer to their experimental measurements, does not improve their rankings in general\cite{wan_ESMACS_2020,wright_ESMACS_2019}. 

To further investigate the structural and energetic precision of Boltz-2 in a hit-to-lead context, a more rigorous, fine-grained ESMACS (FG-ESMACS) protocol was applied to a selected subset consisting of the top 100 compounds from the Boltz-2 predictions for each target. This focused evaluation employed a robust ensemble approach to minimize statistical uncertainty and provide high-resolution validation. For each of the 200 selected protein-ligand complexes, an ensemble of 25 replicas was used in FG-ESMACS. A 10 ns production run was performed for each replica, totalling 250 ns of sampling per compound. Three sets of simulations were performed using distinct initial structures: (1) those generated via docking, (2) Boltz-2 structures with directly added hydrogen atoms, and (3) Boltz-2 structures with hydrogen placements constrained by their SMILES strings (see Section \ref{sec:saturation}).

For the preparation of the docking structures, the selected compounds were first processed using FixpKa\cite{OpenEye} to obtain the correct protonation state at pH 7.4. Subsequently, up to 200 conformers per compound were generated using OMEGA\cite{OpenEye}. The prepared structural library of the compounds was then docked to the protein (PDB IDs: 6W63 for 3CLpro and 4UI5 for TNKS2) using FRED\cite{OpenEye}. The docking poses with the best Chemgauss4 docking scores were used for the ensuing ESMACS simulations. For the structures directly from Boltz-2 prediction, the assigncharges.py script from OpenEye Scientific was used to assign proper protonation states and add explicit hydrogen atoms to represent the charge distribution. To generate Boltz-2 structures with configurations derived from SMILES strings, an in-house Python script was employed to assign hydrogen atoms based on explicit chemical rules rather than simple geometric proximity.

MD simulations were conducted using the NAMD\cite{NAMD} engine and the Amber force field\cite{Amber_FF}. The final binding free energies were calculated using AmberTools\cite{AmberTools}, with results averaged across the 25 replicas to ensure highly converged affinity estimates, providing a reproducible and precise benchmark for evaluating Boltz-2's predictive accuracy. All MD simulations and free energy calculations were performed on Frontier, the world’s first exascale machine at the Oak Ridge National Laboratory in Oak Ridge, Tennessee, United States. Frontier features 37,888 AMD Instinct MI250X with each MI250X containing 2 GPUs. Our FG-ESMACS simulations, comprising 5,000 individual runs, were executed concurrently across 5,000 GPUs (one per run). The entire set of ensemble runs was completed within 112 and 101 minutes for 3CLPro and TNKS2, respectively.

\subsection{Structural and Energetic Evaluation Metrics}

To assess the predictive accuracy of Boltz-2, we employed comparative metrics and statistical analysis focusing on both the geometric fidelity of the predicted complexes and the statistical reliability of the affinity estimations. The structural divergence between the structures predicted by Boltz-2 and those obtained via traditional docking was quantified using two complementary metrics: pairwise root mean square deviations (RMSDs) which were calculated for the main chain of proteins and heavy atoms of ligands, following a rigid-body superposition of the protein backbone to measure global pose displacement; and local distance difference test (LDDT) which is used to assess the preservation of the chemical environment without the bias of global superposition. The LDDT metric evaluates the consistency of interatomic distances between the ligands and binding site residues within a 15 \r{A} radius, providing a high-resolution measure of local interaction fidelity.

The correspondence between Boltz-2 predicted affinities and the binding free energies from ESMACS calculations was evaluated using two primary statistical indicators: the Pearson correlation coefficient (\(r\)) which measures the strength of the linear relationship between the Boltz-2 predictions and the physics-based free energies, and the Spearman’s rank correlation (\(\rho \)) which determines the model's ranking fidelity, assessing its ability to correctly order compounds by potency, a critical factor for virtual screening effectiveness. For the top-100 high-precision subset, we also compared the detailed binding poses and the binding free energies from different initial structures.

\section*{Acknowledgements}
\addcontentsline{toc}{section}{Acknowledgements}

The authors acknowledge the use of resources provided by the Isambard-AI National AI Research Resource (AIRR). Isambard-AI is operated by the University of Bristol and is funded by the UK Government’s Department for Science, Innovation and Technology (DSIT) via UK Research and Innovation, and the Science and Technology Facilities Council [ST/AIRR/I-A-I/1023]. The authors also acknowledge funding support from the DOE INCITE awards for years 2023-2025, which provides access to Frontier at the Oak Ridge Leadership Computing Facility at the Oak Ridge National Laboratory. We thank OpenEye Scientific for academic licenses and support. X.Z. is supported by a China Scholarship Council-UCL Joint Research Scholarship.

\section*{Data availability statement}
\addcontentsline{toc}{section}{Data availability statement}

All datasets generated in this study have been deposited in the Zenodo repository at https://doi.org/10.5281/zenodo.18823303. This includes Boltz-2 predicted structures, binding affinities and binding probabilities for 16,780 compounds (3CLPro) and 21,702 compounds (TNKS2), alongside their corresponding coarse-grained ESMACS binding free energies. Additionally, the repository contains the structural coordinates (Boltz-2 and docking-derived) and high-resolution, fine-grained ESMACS free energy calculations for the top 100 compounds selected from the Boltz-2 predictions.

\section*{Author contributions}
\addcontentsline{toc}{section}{Author contributions}

S.W.: Conceptualization, Data curation, Formal analysis, Methodology, Writing -- original draft, review and editing. X.Z.: Conceptualization, Data curation, Formal analysis, Methodology, Writing -- original draft, review. X.X.: Conceptualization, Writing -- original draft. P.V.C.: Funding acquisition, Conceptualization, Methodology, Supervision, Writing -- original draft, review and editing.

\section*{Competing interests}
\addcontentsline{toc}{section}{Competing interests}

The authors declare no competing interests.

\bibliography{sn-bibliography}

\end{document}